\documentclass[aps,prd,twocolumn,superscriptaddress,longbibliography]{revtex4}
\usepackage[dvipsnames, usenames]{xcolor}
\definecolor{linkcolor}{rgb}{0.0,0.3,0.5}
\usepackage[hypertexnames=false, unicode, colorlinks=true, linkcolor=linkcolor,
citecolor=linkcolor, filecolor=linkcolor,urlcolor=linkcolor,
pdfusetitle]{hyperref}

\usepackage[utf8]{inputenc}
\usepackage{amsmath}
\usepackage{physics}
\usepackage{comment}
\usepackage{bm}
\usepackage{graphicx}
\usepackage{natbib}
\usepackage{mathtools}
\usepackage{xcolor}
\usepackage{ctable}
\usepackage[normalem]{ulem}
\usepackage{multirow}
\usepackage[caption=false]{subfig}
\usepackage[export]{adjustbox}
\usepackage{siunitx}
\usepackage{multirow}

\graphicspath{
	{plots/}
}

\newcommand{\chieff}{\chi_{\mathrm{eff}}}

\usepackage[nomessages]{fp}

\colorlet{RED}{red}

\def\chieff{\chi_{\rm eff}}

\newcommand{\btheta}{\bm{\theta}}

\newcommand{\vabs}[1]{\left\vert\left\vert #1 \right\vert\right\vert}
\newcommand{\mcal}[1]{\mathcal{#1}}
\newcommand{\sk}[1]{}

\allowdisplaybreaks

\begin{document}
	
	\title{Factorized Parameter Estimation for Real-Time Gravitational Wave Inference}
	
	\author{Tousif Islam}
	\email{tislam@umassd.edu}
	\affiliation{Center for Scientific Computing and Visualization Research, University of Massachusetts, Dartmouth, MA 02747, USA}
	\affiliation{Department of Mathematics, University of Massachusetts, Dartmouth, MA 02747, USA}
	\affiliation{Department of Physics, University of Massachusetts, Dartmouth, MA 02747, USA}
	\affiliation{\mbox{Kavli Institute for Theoretical Physics, University of California at Santa Barbara, Santa Barbara, CA 93106, USA}}
	\author{Javier Roulet}
	\affiliation{\mbox{Kavli Institute for Theoretical Physics, University of California at Santa Barbara, Santa Barbara, CA 93106, USA}}
	\affiliation{\mbox{TAPIR, Walter Burke Institute for Theoretical Physics, California Institute of Technology, Pasadena, CA 91125, USA}}
	\author{Tejaswi Venumadhav}
	\affiliation{\mbox{Department of Physics, University of California at Santa Barbara, Santa Barbara, CA 93106, USA}}
	\affiliation{\mbox{International Centre for Theoretical Sciences, Tata Institute of Fundamental Research, Bangalore 560089, India}}

	\date{\today}
	
	\begin{abstract}
		We present a parameter estimation framework for gravitational wave (GW) signals that brings together several ideas to accelerate the inference process. 
		First, we use the relative binning algorithm to evaluate the signal-to-noise-ratio timeseries in each detector for a given choice of intrinsic parameters.
		Second, we decouple the estimation of the intrinsic parameters (such as masses and spins of the components) from that of the extrinsic parameters (such as distance, orientation, and sky location) that describe a binary compact object coalescence.
		We achieve this by semi-analytically marginalizing the posterior distribution over extrinsic parameters without repeatedly evaluating the waveform for a fixed set of intrinsic parameters. 
		Finally, we augment samples of intrinsic parameters with extrinsic parameters drawn from their appropriate conditional distributions.
		We implement the method for binaries with aligned spins, restricted to the quadrupole mode of the signal.
		Using simulated GW signals, we demonstrate that the method produces full eleven-dimensional posteriors that match those from standard Bayesian inference.
		Our framework takes only $\sim \SI{200}{\second}$ to analyze a typical binary-black-hole signal and $\sim \SI{250}{\second}$ to analyze a typical binary-neutron-star signal using one computing core.
		Such real-time and accurate estimation of the binary source properties will greatly aid the interpretation of triggers from gravitational wave searches, as well as the search for possible electromagnetic counterparts. 
		We make the framework publicly available via the GW inference package \texttt{cogwheel}.
		
	\end{abstract}
	
	\maketitle
	
	\section{Introduction}
	The LIGO-Virgo-KAGRA collaborations~\cite{LIGOScientific:2014pky,VIRGO:2014yos,KAGRA:2020cvd} have so far detected close to a hundred gravitational wave (GW) signals from merging binary compact objects~\cite{LIGOScientific:2018mvr,LIGOScientific:2020ibl,LIGOScientific:2021djp,LIGOScientific:2021usb,Olsen:2022pin,pipeline,BBH_O2,1-OGC,2-OGC,2021ApJ...922...76N}. Two of the sources are binary neutron stars (BNS), two are a neutron-star--black-hole (NSBH) binary, and the rest are consistent with being binary black holes (BBHs).
	As the detector sensitivity increases, this number is expected to grow ever faster \cite{KAGRA:2013rdx}. 
	Post detection, subsequent analysis and characterization of the signals is vital for extracting fundamental physics and astrophysics from the events.
	
	The analysis and characterization of GW signals (parameter estimation) is typically performed within the framework of Bayesian inference, and employs stochastic algorithms such as Markov Chain Monte Carlo (MCMC) \cite{Veitch:2014wba,Romero-Shaw:2020owr} or nested sampling~\cite{Speagle_2020} to explore the posterior distribution of the signal parameters. These standard inference techniques have proved effective for most of the cases, but they are computationally expensive to implement. For a typical BBH signal, the straightforward implementation of parameter estimation can take hours to days to run, depending on the nature of the waveform model used and the duration of the signal. For BNS or NSBH signals, run times often extend to weeks. 
	Furthermore, for each event, the analysis is typically repeated using several gravitational waveform models to quantify any systematic biases, and with different initial seeds in the MCMC or nested sampling routines to verify convergence.
	Together, these effects ramp up the run time to days (weeks) for BBH (NSBH and BNS) events.
	
	The detection rate is expected to increase to at least one event per day from the next observing run of the advanced detectors \cite{KAGRA:2013rdx}, and hence, computational cost will be a hurdle for parameter estimation and population inference studies.
	The runtime for parameter estimation is also a significant roadblock in searches for potential electromagnetic counterparts of the GW signals (mostly those that involve neutron stars). 
	Counterpart searches require reliable source classification, and fast and accurate estimation of the sky localization and the distance of the sources, immediately after the detection. 
	
	To address these challenges, several approaches have been developed over the years that employ specialized algorithms using domain-specific knowledge to accelerate the inference process. 
	A significant step was the development of \texttt{Bayestar} \cite{Singer:2015ema}, a rapid sky localization method; under some assumptions, for a fixed set of intrinsic parameters (masses and spins of the compact objects), \texttt{Bayestar} efficiently samples the conditional posterior for extrinsic parameters (such as sky location and distance) that are of interest to electromagnetic follow-up efforts. 
	Using similar ideas, Refs.~\cite{2015PhRvD..92b3002P, Lange:2018pyp, Wofford:2022ykb} developed the highly-parallelized grid-based parameter estimation infrastructure \texttt{RIFT} to sample the posterior in the full parameter space. 
	The crux of the method is to factorize the full parameter space into two blocks: intrinsic and extrinsic parameters, and explore the latter block keeping the former fixed, an idea which we adopt and expand upon in later sections. 
	These codes are much faster than traditional methods, but they either sample only a subset of the parameter space~\cite{Singer:2015ema}, or rely on hardware acceleration~\cite{Lange:2018pyp}. 
	
	Another set of approaches attempt to speed up the evaluation of the likelihood of the data, which is often a rate-limiting step in the sampling process. Ref. \cite{Morisaki:2021ngj} proposed a technique called \textit{multi-banding} to reduce the parameter estimation cost by using coarser frequency resolutions to capture the chirping behavior of the GW signal. 
	It has also been shown that the computational cost of likelihood evaluation, and hence parameter estimation, can be dramatically reduced using relative binning or heterodyning \cite{Cornish:2010kf,Zackay:2018qdy,Dai:2018dca,Leslie:2021ssu,Cornish:2021lje}.
	This method exploits the fact that sampled gravitational waveforms with non-negligible likelihood are similar to each other in the frequency domain, and differ only by fractional perturbations that vary smoothly with frequency. 
	More recently this idea was combined with a sampling scheme tailored to efficiently explore the space of the extrinsic parameters to achieve near real-time parameter inference for aligned-spin quadrupolar BBH signals~\cite{Cornish:2021wxy}.
	Further improvements in computational time have been achieved via a reparametrization of the source properties which removes most of the correlations between parameters, reducing the burden for sampling algorithms~\cite{Roulet:2022kot}.
	A more recent approach has combined the meshfree interpolation methods for the likelihood evaluation with dimension reduction techniques to achieve fast parameter estimation for aligned spin BNS/NSBH signals~\cite{Pathak:2022ktt}. The method, however, only explores a restricted subset of the parameter space.

	Finally, a set of recent efforts have focused on using deep learning methods to speed up GW parameter estimation \cite{Chua:2019wwt,Gabbard:2019rde,Shen:2019vep,Whittaker:2022pkd,Dax:2021tsq,Green:2020dnx,Dax2022}. Deep learning methods involve training a neural network with a large number of underlying hyper-parameters in addition to the parameters describing compact binary coalescence, using injected signals as training data. 
	These methods have recently achieved remarkable speed and accuracy. Currently, however, they are applicable only to short signals from massive BBH mergers.
	The training of the neural network also incurs significant offline computational cost and requires specialized hardware.
	
	We unite salient elements from the previous disparate approaches into a framework to perform real-time GW parameter estimation. 
	We make the framework publicly available via the GW inference package \texttt{cogwheel} \cite{cogwheel}.
	We first decouple the estimation of intrinsic parameters (such as masses and spins of the binary) and the extrinsic parameters (such as distance, sky localization and orientation of the binary), in a manner that is akin to the approach presented in Ref.~\cite{2015PhRvD..92b3002P}, with some additional improvements.
	We draw samples from the posterior for the intrinsic parameters of the mergers, which has been marginalized over the extrinsic parameters. 
	We perform this marginalization semi-analytically; to be specific, we analytically marginalize over the distance and phase of the signal, and use a combination of simple Monte Carlo sampling and importance sampling over the other extrinsic parameters (this method has been previously used in a search pipeline as part of a refined detection statistic \cite{Olsen:2022pin}). 
	The sampling of the intrinsic parameters further exploits the relative binning method proposed in Ref. \cite{Zackay:2018qdy} with some modifications.
	Once we have samples of the intrinsic parameters, we complete them in postprocessing by appending extrinsic parameters drawn from the posterior distribution conditioned on the intrinsic parameters.
	We find that our method can provide rapid estimation of the source properties without compromising accuracy for all compact binary signals (i.e. BBH, BNS or NSBH) with aligned-spin configurations and quadrupole mode of radiation. 
	The runtime depends mostly on the signal-to-noise ratio.
	For typical GW signals, it can generate results in $\sim \SI{200}{\second}$ using one computing core irrespective of the nature of the binary. For GW170817, the loudest BNS signals detected so far, our framework takes $\sim \SI{20}{\minute}$ to analyze using one computing core.
	
	The rest of the paper is organized in the following way. Section \ref{sec:methods} describes our parameter estimation framework while \ref{sec:result} presents our results. We finally discuss the limitations of our method and propose future improvement in Section \ref{sec:discussion}.
	
	\section{Parameter Estimation Methodology}
	\label{sec:methods}
	In this section, we summarize Bayesian inference methods (Sec. \ref{subsec:bayes}) and present our strategy for rapid parameter estimation of the GW source properties (Sec. \ref{subsec:rapid_pe}).
	
	\subsection{Basics of Bayesian Inference}
	\label{subsec:bayes}
	In the Bayesian framework, for a set of the GW source parameters $\theta$, the probability distribution $p(\theta|d)$ is computed via Bayes' theorem:
	\begin{equation}
		p(\btheta|d) = \frac{\pi(\btheta) \mathcal{L}(d|\btheta)}{\mathcal{Z}(d)}
	\end{equation}
	where $d(t)$ represents time-domain strain data which is assumed to be a sum of the true signal $h(t; \btheta)$ and stationary Gaussian noise $n(t)$:
	\begin{equation}
		d(t) = h(t; \btheta) + n(t).
	\end{equation}
	The quantity $\mathcal{L}(d|\btheta)$ is the \textit{likelihood} of $d(t)$ given a set of GW source parameters $\btheta$; $\pi(\btheta)$ is the \textit{prior} probability of $\btheta$ (which is determined based on our knowledge or belief on $\btheta$); and $\mathcal{Z}(d)$ is the \textit{evidence} (marginalized likelihood) of $d(t)$. 
	
	Under the approximation that the noise is stationary and Gaussian, the likelihood takes the following form: 
	\begin{equation}
		\log \mathcal{L}(d|\btheta) = - \frac{1}{2} \sum_{k \in \rm det} 4 \int_{0}^{\infty} \frac{|\tilde{d}_k(f)-\tilde{h}_k(f;\btheta)|^2}{S_k(f)} {\rm d}f,
		\label{eq:likelihood}
	\end{equation}
	where $\tilde{d}_k(f)$, $\tilde{h}_k(f)$ and $S_k(f)$ are the frequency-domain strain, strain model, one-sided power spectrum in the $k$th detector. 
	
	Under the assumption that the binaries are quasi-circular and non-precessing, the vector $\btheta=(\mcal{I}, \mcal{E})$ is a set of at least eleven parameters that completely characterize a binary compact object GW signal in general relativity. It is consists of  at least four intrinsic parameters $\mcal{I}$ and seven extrinsic parameters $\mcal{E} := \{\mcal{E}_m, \mcal{E}_d\}$. 
	The vector $\mcal{E}_m := \{\iota, \varphi_c\}$ denotes the direction of radiation in the source frame: $\iota$ is the inclination angle between the orbital angular momentum of the binary and line-of-sight to the observer whereas $\varphi_c$ is the orbital phase angle at coalescence. 
	Extrinsic parameters $\mcal{E}_d := \{\alpha, \delta, \psi,  \tau_c, D\}$ describe the location of the merger relative to the detector.
	The luminosity distance of the binary is denoted by $D$.
	Right ascension $\alpha$ and declination $\delta$ parameterize sky localization, $\psi$ is the polarization angle and $t_c$ is the time at coalescence. 
    The vector $\mathcal{I}:=\{m_1,m_2,\chi_1,\chi_2\}$ contains the intrinsic parameters that describe the binary: the component masses $m_1$ and $m_2$ (with $m_1>m_2$) and dimensionless spin magnitudes $\chi_1$ and $\chi_2$. 
	The parameter space will also include the tidal deformability parameters if the compact objects are neutron stars.
	
	The choice of the model for $h(t;\btheta)$ (typically referred to as \textit{waveform model}) then defines a particular signal hypothesis. Most of the waveform models covering the  inspiral, merger and ringdown (IMR) phases of the binary coalescence can typically be classified into four categories: numerical relativity (NR) based surrogate models~\cite{Field:2013cfa,Blackman:2015pia,Blackman:2017pcm,Blackman:2017dfb,Varma:2018mmi,Varma:2019csw,Islam:2021mha}, effective-one-body models~\cite{bohe2017improved,cotesta2018enriching,cotesta2020frequency,pan2014inspiral,babak2017validating}, phenomenological models~\cite{husa2016frequency,khan2016frequency,london2018first,khan2019phenomenological} and NR informed perturbation theory based surrogate models \cite{Rifat:2019ltp,Islam:2022laz}. All of these models typically provide a select number of higher harmonics apart from the dominant quadrupolar $(\ell,m)=(2,\pm2)$ mode of gravitational radiation. We restrict ourselves to only $(\ell,m)=(2,\pm2)$, aligned-spin waveforms in this work. This simplifies the dependence of the waveform on inclination and orbital phase.
	
	Once the likelihood $\mathcal{L}(d|\btheta)$, priors $\pi(\btheta)$ and a waveform model are chosen, samples from the posterior distribution $p(\btheta|d)$ are generated through a sampling algorithm: either by MCMC \cite{Veitch:2014wba,Romero-Shaw:2020owr} or by nested sampling~\cite{Speagle_2020}. This is typically done by specific GW parameter estimation packages such as \texttt{bilby}~\cite{Ashton:2018jfp,Smith:2019ucc}, \texttt{lalinference}~\cite{Veitch:2014wba} or \texttt{cogwheel}~\cite{Roulet:2022kot} with samplers such as \texttt{dynesty}~\cite{Speagle_2020}, \texttt{PyMultiNest}~\cite{Feroz:2008xx} or \texttt{emcee}~\cite{Foreman_Mackey_2013}.

	\subsection{Factorized Parameter Estimation Framework}
	\label{subsec:rapid_pe}
	Sampling over a multi-dimensional parameter space (for our case, eleven dimensional) is computationally challenging. For some waveform models with richer phenomenology, the likelihood function $\mathcal{L}(d|\btheta)$ can be costly to evaluate. The large volume of the phase space may also make the sampling inefficient or slow. 
	Therefore, the dimensionality of the sampling space is often reduced by analytically marginalizing over certain binary parameters (typically luminosity distance and the time at coalescence). This helps in both speeding up computation and sampling convergence. Below, we present a fast parameter estimation framework within \texttt{cogwheel} based on semi-analytical marginalization over all extrinsic parameters.

	\subsubsection{Choice of priors}
	\label{subsubsec:prior}
	We choose uniform priors for the detector frame masses for the two compact objects $m_{1,\rm det}$ and $m_{2,\rm det}$ and uniform prior for the effective inspiral spin $\chi_{\rm eff}$.
	The prior on the luminosity distance is taken as $p(d_L)\propto d_L^2$. For the inclination angle $\iota$, we choose $p(\iota)\propto \sin\iota$ where $0\le\iota\le \pi$. Priors on the orbital phase $\varphi_c$, time at coalescence $t_c$, sky localization angles \{$\alpha,\sin\delta$\} and polarization angle $\psi$ are taken to be uniform.
	
	\subsubsection{Marginalized likelihood}
	\label{subsubsec:marg}
	The log-likelihood for the data, given a set of parameters $\{\mcal{I},\mcal{E}\}$, is
	\begin{align}
		2 \log \mathcal{L}(d | \mcal{I}, \mcal{E})
		&= - \sum_{k \in {\rm detectors}} \langle d_k - h_k \vert d_k - h_k \rangle_k,
		\label{likelihood_simple}
	\end{align}
	where the inner product $\langle . \vert . \rangle$ is weighted by $S_k^{-1}(f)$.
	Eq.~(\ref{likelihood_simple}) can be re-written as:
	\begin{align}
		2 \ln \mathcal{L}(d | \mcal{I}, \mcal{E})
		=&- \sum_k \langle d_k \vert d_k \rangle_k+\frac{\abs{\mathbf{z}_0 \cdot \mathbf{Z}}^2}{\vabs{\mathbf{z}_0}^2} \notag \\
		&- \vabs{\mathbf{z}_0}^2 \abs{Y - \frac{\mathbf{z}_0 \cdot \mathbf{Z}}{\vabs{\mathbf{z}_0}^2}}^2. 
		\label{eq:lnL_simpl}
	\end{align}
	Here, $\mathbf{Z}(\mcal{I}, \tau_c, \alpha, \delta)$ is a vector of complex overlaps between template and data in detectors:
	\begin{align}
		\mathbf{Z}(\mcal{I}, \tau_c, \alpha, \delta) & = \begin{pmatrix}
			\vdots \\
			Z_k(\mcal{I}, \tau_k(\tau_c, \alpha, \delta)) \\
			\vdots \\
		\end{pmatrix}. \label{eq:complexz}
	\end{align}
	We also define a vector of complex overlaps in detectors if the data were composed of just the waveform (without any noise).
	\begin{align}
		\mathbf{z}(\mcal{I}, \mcal{E}_m, \mcal{E}_d\setminus\tau_c) & = \begin{pmatrix}
			\vdots \\
			z_k(\mcal{I}, \mcal{E}_m, \mcal{E}_d\setminus\tau_c) \\
			\vdots \\
		\end{pmatrix}.
	\end{align}
	Here, $Y = (D_0/D) e^{-2 i \varphi_c}$ where $D_0$ is a fiducial reference distance. The symbol $\mcal{E}_d\setminus\tau_c$ indicates a quantity that depends on the parameters in $\mcal{E}_d$ apart from the merger time $\tau_c$.
	We further define $\mathbf{z}_0(\mcal{I}, \mcal{E}_m\setminus\varphi_c, \mcal{E}_d\setminus\{D, \tau_c \})$ which is a vector of complex overlaps that would be attained for a signal from a merger with phase $\varphi_c=0$ and at the fiducial reference distance $D_0$. 
	
	We now multiply the likelihood $\mathcal{L}(d|\mcal{I}, \mathcal{E})$ (Eq.~\eqref{eq:lnL_simpl}) with a prior $\Pi$ over all seven extrinsic parameters $\mcal{E}$, and integrate. This gives the probability that the data contains a GW signal with known intrinsic parameters $\mcal{I}$. 
	\begin{align}
		p(d \vert \mcal{I}) & = \int d\Pi (\mcal{E}) \mathcal{L}(\mcal{I}, \mcal{E}).
		\label{eq:marg_likelihood}
	\end{align}
	This is interpreted as the likelihood marginalized over extrinsic parameters.
	We now substitute Eq.(\ref{eq:marg_likelihood}) into Eq.(\ref{eq:lnL_simpl}). With a bit of a bit of algebra, for quasi-circular quadrupolar aligned-spin binaries, we can rewrite the marginalized likelihood Eq.~(\ref{eq:marg_likelihood}) as (see Appendix D in Ref.~\cite{Olsen:2022pin} for the details)
	\begin{align}
		&\mathcal{L}_{\rm marg}(d|\mcal{I}) \notag \\
		&= \frac{4\pi \Pi_0 D_0^3}{T} \prod_{k \in \, {\rm detectors}} \int {\rm d}\tau_k \, \exp{ \left( \frac12 \abs{Z_k(\tau_k)}^2 \right)}  \notag \\
		& \times  \int  \frac{{\rm d}\alpha}{2 \pi} \frac{{\rm d}\delta \cos{\delta}}{2} {\rm d}\tau_c \, \dd(\tau_k - \tau_k(\tau_c, \alpha, \delta))\, \notag \\
		&  \times \int \frac{{\rm d}\mu}{2} \frac{{\rm d}\psi}{2\pi} \, \exp{ \left\{ - \frac12 \left( \vabs{\mathbf{Z}}^2 - \abs{\hat{\mathbf{z}}_0\left( \mu, \alpha, \delta, \psi \right) \cdot \mathbf{Z} }^2 \right) \right\} }  \notag \\
		&  \times \vabs{\mathbf{z}_0}^3 g\left( \vert \hat{\mathbf{z}}_0 \cdot \mathbf{Z} \vert \right), \label{eq:ldphi2}
	\end{align}
	where $\mu=\cos\iota$, $\tau_k$ is the arrival time in each detector, $T$ is the time window which contains the time at coalescence $\tau_c$, $\dd$ denotes Dirac-delta functions and $\hat{\mathbf{z}}_0=\frac{\mathbf{z}_0}{\vabs{\mathbf{z}_0}}$.
	The functional form of $g\left( \vert \hat{\mathbf{z}}_0 \cdot \mathbf{Z} \vert \right)$ is known (see Appendix D in Ref. \cite{Olsen:2022pin} for the details). 
	
	\subsubsection{Relative Binning}
	The marginalization over extrinsic parameters takes as inputs the matched-filtering timeseries $\langle d \mid h(t) \rangle$ of a template with the data and its normalization $\langle h \mid h \rangle$, for each detector. We compute both of these quantities using the relative-binning method \cite{Zackay:2018qdy}, with the following modification. Relative binning requires that the waveforms of interest resemble a reference waveform, while in practice we have to compute the matched-filtering timeseries over a $\sim \SI{100}{\milli\second}$ window of time around the event, much larger than the autocorrelation length of the waveform. To ensure that any waveform remains similar to the reference over all arrival times, we compute summary data for a large set of arrival times at each detector.
	Unlike the Fast Fourier Transform algorithm, this enables us to obtain the matched-filtering timeseries from irregularly sampled waveforms.
	
	\subsubsection{Estimation of the Intrinsic Parameters}
	To sample over the intrinsic parameters (i.e. the masses $m_1$ and $m_2$ and the spins $\chi_1$ and $\chi_2$), we choose $\mathcal{L}_{\rm marg}(d|\mcal{I})$ to be our marginalized likelihood.
	Sampling is done using relative binning as implemented in \texttt{cogwheel} with \texttt{PyMultiNest} which is one of the state-of-the-art nested samplers. For BBHs, we choose \texttt{IMRPhenomXAS}~\cite{Pratten:2020fqn}, a frequency-domain phenomenological model for aligned-spin BBHs and, for BNS, we use \texttt{IMRPhenomD\_NRTidalv2}~\cite{Dietrich:2018uni}, a state-of-the-art frequency domain phenomenological model that incorporates tidal effects.
	
	We formulate a Monte-Carlo method to compute the marginalized likelihood.
	We discretize the arrival times in each detector, $\tau_k$, on a time grid which is fine enough to capture any structure in the $Z(\tau_k)$ time-series. We use a sampling frequency $1/\Delta \tau_k=\SI{4096}{\hertz}$.
	The sky, parameterized by $\{\alpha, \delta\}$, is also discretized into $n_{ \alpha}\times n_{\delta}$ cells of equal area. 
	The values for $n_{\alpha}$ and $n_{\delta}$ are chosen to provide a high enough angular resolution for the sky.
	With these samples we construct a dictionary, mapping discretized time-of-arrival differences between the detectors in the network to the sky location samples corresponding to those time delays.
	Finally, for the polarization $\psi$ and cosine of inclination angle $\mu$, we generate $n_{\rm MC}$ Monte Carlo points from their priors. We use $n_{\rm \rm MC}=\num{10000}$ to ensure a dense sampling.
	
	We assign a sky location and time of arrival to each Monte Carlo sample as follows. We select $n_{\rm MC}$ tuples of arrival times in all detectors $(\cdots, \tau_k, \cdots)$ with each component $\tau_k$ picked according to a probability $\sim \exp{ \left( \frac12 \abs{Z_k(\tau_k)}^2 \right)}$, which acts as an importance sampling proposal. Tuples $(\cdots, \tau_k, \cdots)$ that do not obey physically allowed time delays between detectors are discarded, the rest of the tuples are used to pick sky location samples using the dictionary.
	The integration over the distance $D$ and orbital phase are done analytically, while the integration over sky angles $\{\alpha,\delta\}$, polarization $\psi$ and inclination $\iota$ are done numerically via Monte Carlo. For each tuple $(\tau_{c,j}, \alpha_j, \delta_j, \psi_j, \mu_j)$, we also record their contribution to the integral (i.e.\ the `weights') in Eq.~\eqref{eq:ldphi2}.

	\subsubsection{Estimation of right ascension, declination, polarization and inclination angles }
	Once the intrinsic parameters are obtained, for each tuple of $(m_{1,j},m_{2,j},\chi_{1,j},\chi_{2,j})$, we re-construct all possible $(\alpha_j,\delta_j,\psi_j,\iota_j)$ tuples along with their weights. We then randomly select one of the $(\alpha_j,\delta_j,\psi_j,\iota_j)$ tuples according to their weights  and assign them to $(m_{1,j},m_{2,j},\chi_{1,j},\chi_{2,j})$. During this process, we also save the corresponding overlap tuple $(\vabs{\mathbf{z}_{0,j}}^2, \mathbf{z}_{0,j} \cdot \mathbf{Z}_j)$ and merger time $\tau_{c,j}$, which are useful for sampling the time, distance and phase later on.
	
	\subsubsection{Estimation of the distance}
	For fixed values of $(m_{1,j},m_{2,j},\chi_{1,j},\chi_{2,j},\alpha_j,\delta_j,\psi_j,\iota_j,\allowbreak t_{c,j})$ and their corresponding overlap tuple $(\vabs{\mathbf{z}_{0,j}}^2, \mathbf{z}_{0,j} \cdot \mathbf{Z}_j)$, distance $D$ and phase $\varphi_c$ are sampled from the joint probability distribution of $\{D,\varphi_c\}$ (see Appendix D in Ref.~\cite{Olsen:2022pin} for the details):
	\begin{multline}
		P(D,\varphi_c)\\
		\propto  D^2 \Pi(D) \frac{1}{2\pi} \exp{\left( - \frac{\vabs{\mathbf{z}_{0,j}}^2}{2} \abs{ \frac{D_0}{D} e^{-2 i \varphi_c} - \frac{\mathbf{z}_{0,j} \cdot \mathbf{Z}_j}{\vabs{\mathbf{z}_{0,j}}^2}}^2 \right)}.
		\label{eq:Idphi}
	\end{multline}
	We choose the fiducial distance to be $D_0=\SI{1}{Mpc}$ and construct a dummy variable $y=D_0/D$. We then obtain the posterior distribution for $y(D)$ by marginalizing $P(D,\varphi_c)$ over the phase:
	\begin{equation}
	\begin{split}
		P(y) & \propto \int P(y(D),\varphi_c) \frac{d\varphi_c}{2\pi}  \\
		& \propto \frac{1}{y^4} \exp{\left( - \frac{\vabs{\mathbf{z}_{0,j}}^2}{2} \left[ y - \frac{\vert \mathbf{z}_{0,j} \cdot \mathbf{Z}_j \vert}{\vabs{\mathbf{z}_{0,j}}^2} \right]^2 \right)} \\
		&\quad \times I_0 \left( y \vert \mathbf{z}_{0,j} \cdot \mathbf{Z}_j \vert \right),
		\label{eq:dist_pick}
	\end{split}
	\end{equation}
	where $I_0$ is the modified Bessel function of the first kind. 
	The structure of Eq.~\eqref{eq:dist_pick} suggests that the peak for $P(y)$ is near
	$y_{\rm peak} = \vert \mathbf{z}_{0,j} \cdot \mathbf{Z}_j \vert / \vabs{\mathbf{z}_{0,j}}^2$ and has a width $\sigma_{y}=1/\vabs{\mathbf{z}_{0,j}}^2$. 
	We construct a grid of 1000 uniformly spaced points $\{y_k\}$ in the interval $y_k \in [y_{\rm min}, y_{\rm max}]$ where $y_{\rm min}={\max}(10^{-5}, y_{\rm peak}-7\sigma_{y})$ (to make sure that $y_{\rm min}$ is never zero or negative) and $y_{\rm max} = y_{\rm peak} + 7 \sigma_{y}$.
	We extend the grid by $7\sigma_y$ away from $y_{\rm mean}$ in both sides to make sure all points with non-negligible probabilities are included.
	We then use inverse-transform sampling to draw a value of $y$: we obtain the normalized cumulative distribution function 
	\begin{equation}
		F(y) = \frac {\sum_{y_{j'} \le y} P(y_{j'})} {\sum_{y_{j'}\le y_{\rm max}} P(y_{j'})},
	\end{equation}
	and choose a point $y_j$ whose value of $F$ matches a value drawn uniformly between 0 and 1. The luminosity distance is then simply given by $D_j=D_0 / y_j$.
	
	\subsubsection{Estimation of the orbital phase}
	Next, we employ a similar method to sample the orbital phase $\varphi_c$ from the following distribution:
	\begin{align}
		P(\varphi_c) 
		&  \propto \exp{\left[ {\rm Re} \left( y_j^\ast \mathbf{z}_{0,j} \cdot \mathbf{Z}_j \exp{(- 2 i \varphi_c)} \right) \right]}.
		\label{eq:phase_pick}
	\end{align}
	At this step, all necessary components (i.e.\ $y_j$ and $\mathbf{z}_{0,j} \cdot \mathbf{Z}_j$) are known. If we inspect Eq.~(\ref{eq:phase_pick}), we obtain $\varphi_{c, \rm peak}=\rm arg(\mathbf{z}_{0,j} \cdot \mathbf{Z}_j)/2$ and define 
	\begin{equation}
		\Delta \varphi_c = \varphi_c - \varphi_{\rm c,peak}.
	\end{equation}
	The posterior distribution for $\Delta \varphi_c$ simply becomes:
	\begin{align}
		P(\Delta \varphi_c) 
		&  \propto \exp{\left[ y_j \vert \mathbf{z}_{0,j} \cdot \mathbf{Z}_j \vert \cos{(2 \Delta \varphi_c)} \right]},
		\label{eq:dphase_pick}
	\end{align}
	which has identical peaks at $\Delta \varphi_{c,\rm peak}=0, \pi$ each with a standard deviation $\sigma_{\Delta \varphi_{c}}=(\vert  \mathbf{z}_{0,j} \cdot \mathbf{Z}_j \vert y_j)^{-1/2}$.
	We now construct a grid of 1000 uniformly spaced points $\{\Delta \varphi_{c,k}\}$ in the interval $\Delta \varphi_{c,j} \in [-\Delta \varphi_{c,\rm max}, \Delta \varphi_{c,\rm max}]$ where $\Delta \varphi_{c,\rm max}={\min}(7\sigma_{\Delta \varphi_{c}},\pi/2)$. From these we draw a value of $\Delta \varphi_{c,j}$ with probabilities given by Eq.~\eqref{eq:dphase_pick}. 
	Finally, we account for degeneracy in $(\ell,m)=(2,\pm2)$ modes under a orbital phase rotation of $\pi$:
	\begin{equation}
		\varphi_{c,j} = \varphi_{c, \rm mean} + \Delta \varphi_{c,j} + \gamma_j \pi,
	\end{equation}
	where $\gamma_j$ is chosen randomly from $\{0,1\}$.
	This completes the estimation of the tuple consisting all binary source parameters: 
	$
 (m_{1,j}, \allowbreak
  m_{2,j}, \allowbreak
  \chi_{1,j}, \allowbreak
  \chi_{2,j}, \allowbreak
  \alpha_j, \allowbreak
  \delta_j, \allowbreak
  \psi_j, \allowbreak
  \iota_j, \allowbreak
  D_j, \allowbreak
  \varphi_{c,j}, \allowbreak
  t_{c,j})$. 
	
	\begin{table}
		\caption{Source properties of the simulated BBH events used in this work.}\footnote{Symbols:  $\mathcal{M}^\mathrm{det}$ : Detector-frame chirp mass; $q$: Mass ratio;  $\chi_\mathrm{eff}$: Effective inspiral spin;  $D$: Luminosity distance; $\iota$: Inclination; $\alpha$: right ascension; $\delta$: declination, $\psi$: polarization.}
		\begin{ruledtabular}
			\begin{tabular}{c | c | c | c | c}
				&\texttt{Event-1} &\texttt{Event-2} &\texttt{Event-3} &\texttt{Event-4}\\
				\hline
				$\mathcal{M}^\mathrm{det}/ M_\odot$&28.72 &17.41 &26.76 &20.51 \\
				$q$ &1.0 &1.0 &0.5 &0.33\\
				$\chi_\mathrm{eff}$ &0.0 &0.25 &0.267 &0.0\\
				\rule{0pt}{4ex}%
				$D/\mathrm{Mpc}$ &1000 &750 &1000 &500\\
				$\iota$ &$\pi/4$ &$\pi/4$ &$\pi/4$ &$\pi/4$\\
				$\varphi_c$ &$\pi/5$ &$\pi/5$ &$\pi/5$ &$\pi/5$\\	
				\rule{0pt}{4ex}
				$\alpha$  &2.4 &2.4 &2.4 &2.4\\
				$\delta$ &0.15 &0.15 &0.15 &0.15\\
				$\psi$ &0.5 &0.5 &0.5 &0.5\\
			\end{tabular}
		\end{ruledtabular}
		\label{tab:Injections}
	\end{table}
	
	\section{Result}
	\label{sec:result}
	In this section, we demonstrate the efficacy of our parameter estimation method using an injection study and provide an estimate of the computational costs.
	
	\subsection{Analysis of the simulated BBH events}
	We create a set of four representative simulated aligned-spin BBH signals with only $(\ell,m)=(2,\pm2)$ mode of radiation using the \texttt{IMRPhenomXAS} waveform model.
	These signals are then injected in the LIGO-Virgo O3a noise data. 
	We choose varying masses, spins and distances for these binaries to explore different regions of the binary parameter space.
	The source properties of the simulated events are shown in Table \ref{tab:Injections}.
	
\begin{figure*}[thb]
		\includegraphics[scale=0.75]{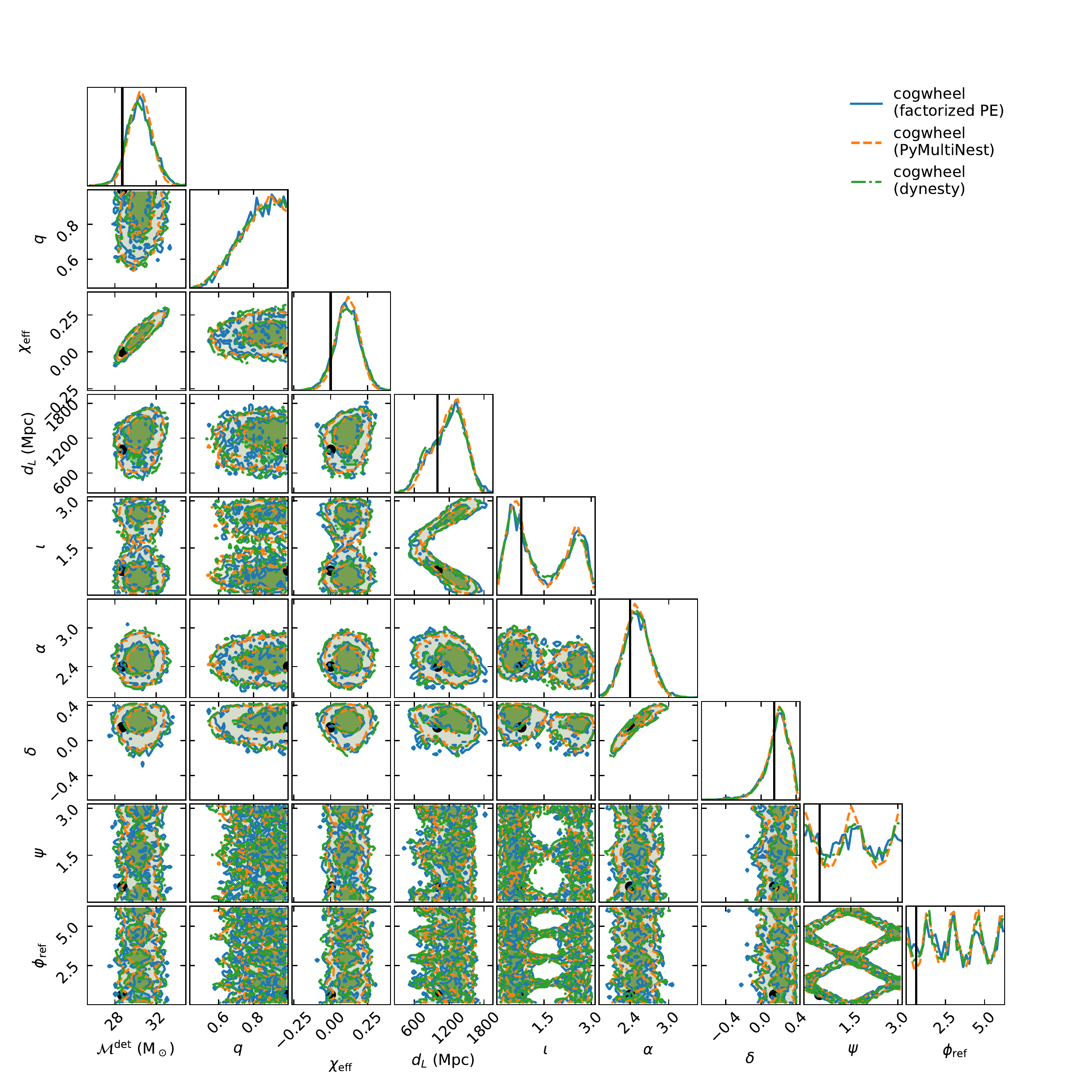}
		\caption{Comparison between parameter estimation results from standard Bayesian frameworks (with \texttt{dynesty} and \texttt{PyMultiNest} samplers respectively) and result from the factorized parameter estimation framework for an injected signal (\texttt{Event-1} in Table \ref{tab:Injections}) with detector frame total mass $M=66 M_{\odot}$, mass ratio $q=1$, $d_L=1000$ Mpc. Injected values are shown in black. Contours enclose 50\% and 90\% of the posterior probability.
		}
		\label{Fig:injected_study}
	\end{figure*}

	We first analyze \texttt{Event-1}, whose source properties are largely consistent with GW150914 (with detector frame total mass $M=66 M_{\odot}$, mass ratio $q=1$, spins $\chi_1=0$, $\chi_2=0$, luminosity distance $D=1000$ Mpc and inclination angle $\iota=\pi/4$). 
	We estimate the posterior density functions for all GW-source parameters with the factorized parameter estimation framework. 
	For the analysis, we set the number of points in $\alpha$ and $\delta$ grids as: $n_{\alpha}=n_{\delta}=2000$. The number of live points for \texttt{PyMultiNest} is chosen such a way that it is enough to explore the four dimensional parameter space defined by the masses and the spins. We set $\texttt{nlive}=512$.
	For comparison, we also perform the parameter estimation with the standard Bayesian framework in \texttt{cogwheel} with two different samplers: \texttt{dynesty} and \texttt{PyMultiNest}.
	For the standard Bayesian analyses, where the sampler explores \textit{eleven} dimensional parameter space of the binary, the number of live points have been increased to $\texttt{nlive}=2048$. 
	In Fig.~\ref{Fig:injected_study}, we show a comparison between posteriors inferred using the factorized parameter estimation framework (solid blue), and using standard method sampling over eleven parameters with the \texttt{PyMultiNest} (dashed orange) and \texttt{dynesty} (dashed-dotted green) samplers respectively. Injected values are shown as black solid lines (in one dimensional posteriors) and black dots (in two dimensional contours). We demonstrate that our rapid parameter estimation framework results agree well with the standard Bayesian results.
	
	\begin{table}
		\caption{Summary of the JS divergence values (in nats) between the one-dimensional marginalized PDFs for \texttt{Event-1} obtained using the factorized rapid parameter estimation framework, standard Bayesian analysis with \texttt{dynesty} sampler and the standard Bayesian analysis with \texttt{PyMultiNest} sampler, respectively.}
		\footnote{Symbols:  $\mathcal{M}^\mathrm{det}/ M_\odot$ : Detector-frame chirp mass; $q$: Mass ratio;  $\chi_\mathrm{eff}$: Effective inspiral spin parameter;  $D/\mathrm{Mpc}$: Luminosity distance; $\iota$: Inclination angle; $\alpha$: right ascension; $\delta$: declination, $\psi$: polarization.}
		\begin{ruledtabular}
			\begin{tabular}{c | c | c}
				&JS-Divergnece &JS-Divergence\\
				&between &between\\ 			
				&factorized PE \& &factorized PE \&\\
				&\texttt{cogwheel} (\texttt{dynesty}) &\texttt{cogwheel} (\texttt{PyMultiNest})\\
				\hline
				$\mathcal{M}^\mathrm{det}/ M_\odot$ & 0.030 &0.081 \\
				$q$ &0.017 &0.021\\
				$\chi_\mathrm{eff}$ &0.029 &0.078\\
				\rule{0pt}{4ex}%
				$D/\mathrm{Mpc}$ &0.051 &0.075\\
				$\iota$ &0.017 &0.046\\
				$\varphi_c$ &0.021 &0.029\\	
				\rule{0pt}{4ex}
				$\alpha$  &0.034 &0.062\\
				$\delta$ &0.040 &0.064\\
				$\psi$ &0.023 &0.051\\
			\end{tabular}
		\end{ruledtabular}
		\label{tab:JSdivergence}
	\end{table}

	\begin{figure*}[thb]
		\includegraphics[scale=0.75]{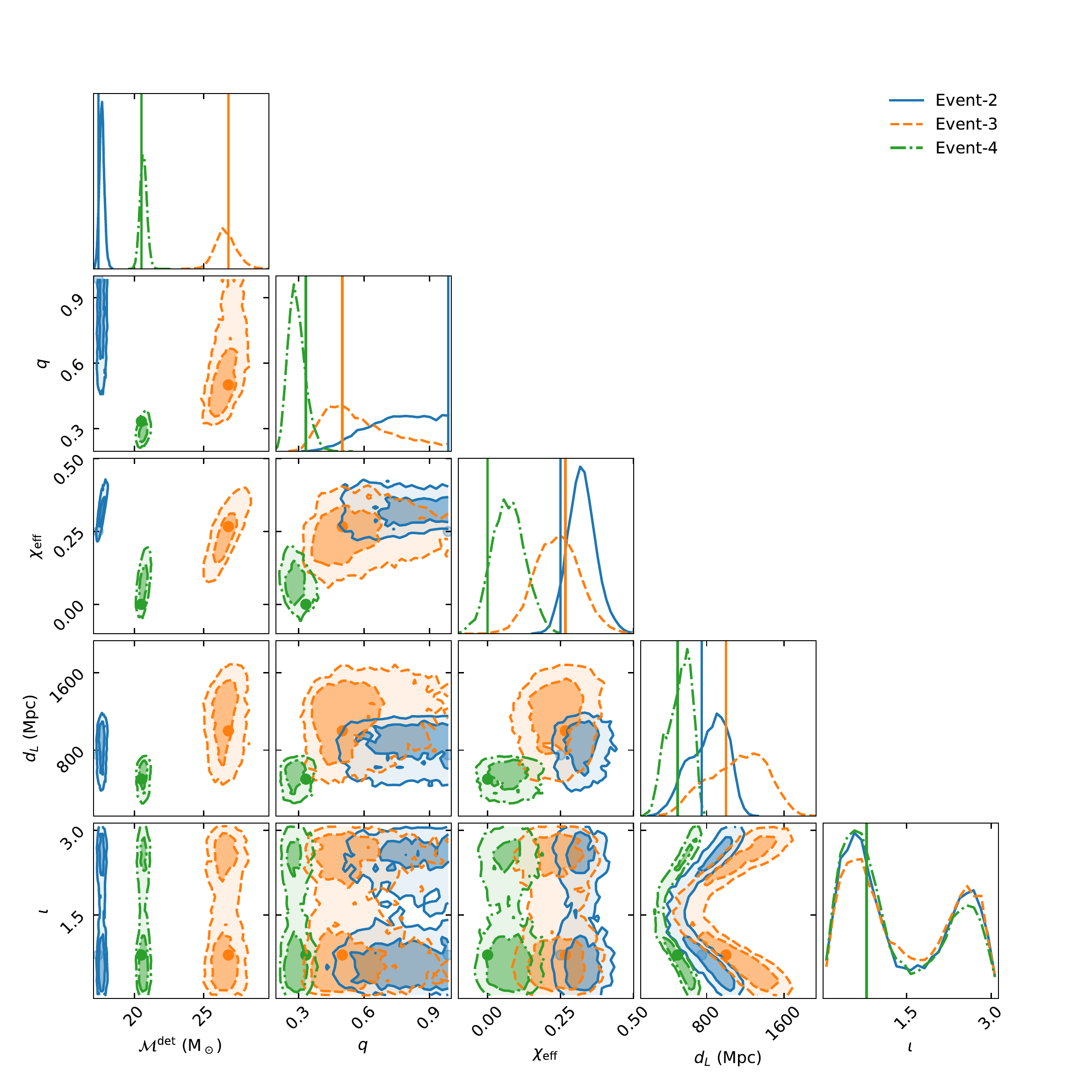}
		\caption{Posteriors obtained by analyzing \texttt{Event-2} (blue), \texttt{Event-3} (orange) and \texttt{Event-4} (green) using the factorized parameter estimation framework respectively. Vertical lines (dot) indicates true values for 1d marginalized posteriors (2d contours). Contours enclose 50\% and 90\% of the posterior probability.
		}
		\label{Fig:Event_2_3_4}
	\end{figure*}
	We then quantify the difference between the one-dimensional marginalized posteriors obtained in these three different cases using the Jensen-Shannon (JS) divergence values~\cite{Js61115}.
	The JS divergence is a general symmetrized extension of the Kullback-Leibler divergence~\cite{KLdivergence}. The JS divergence between two probability vectors $p(x)$ and $q(x)$ is defined as:
	\begin{eqnarray}
		D_{\rm JS}(p,q) = \sqrt{\frac{D_{\rm KL}(p||m) + D_{\rm KL}(q||m)}{2}},
	\end{eqnarray}
	where $m(x)$ is the point-wise mean of $p(x)$ and $q(x)$ and $D_{\rm KL}$ is Kullback-Leibler divergence:
	\begin{equation}
		D_{\rm KL}(p||q) = \int p(x) \log\frac{p(x)}{q(x)} dx.
	\end{equation}
	JS divergence values of 0 signify that the posteriors are identical while a JS divergence value of $\log 2$ nat would mean the posterior distributions have no statistical overlap at all. Values above 0.15 nats are sometimes considered to reflect non-negligible bias~\cite{LIGOScientific:2018mvr}. In Table \ref{tab:JSdivergence}, we summarize the JS divergence values (computed using \texttt{scipy.spatial.distance.jensenshannon} module) between posterior distributions for a select set of binary parameters. We find that the JS divergence values between factorized parameter estimation result and standard Bayesian \texttt{cogwheel} results are always less than 0.1 nats suggesting excellent agreement. We also find that the factorized parameter estimation result agrees more closely with the fully Bayesian results obtained with \texttt{dynesty} sampler than the one obtained with \texttt{PyMultiNest}.
	
	\begin{table}[h!]
		\centering
		\begin{tabular}{lr}
			\toprule
			PE framework                 & Time taken        \\
			\hline
			\multirow{1}{*}{\texttt{cogwheel} (\texttt{PyMultiNest})}   & \SI{1782}{\second} \\					
			\\						
			\hline
			\multirow{1}{*}{\texttt{cogwheel} (\texttt{dynesty})}   & \SI{4226}{\second}\\
			\\
			\hline
			\multirow{3}{*}{\texttt{cogwheel} (\texttt{factorized pe})}   & \\\\
			total time & \SI{185}{\second}     \\
			finding maximum likelihood & \SI{36}{\second}     \\
			sampling parameters & \SI{149}{\second}     \\
			\botrule	
		\end{tabular}
		\caption{Summary of the time taken for different parameter estimation frameworks using one core in analyzing the simulated BBH GW signal \texttt{Event-1} in Table \ref{tab:Injections}.}
		\label{Tab:sampling_time}
	\end{table}
	We measure the time taken for the rapid parameter estimation framework as well as the standard Bayesian framework with these two different samplers in analyzing \texttt{Event-1}.
	In Table \ref{Tab:sampling_time}, we summarize the result of our timing experiment.
	We find that the standard Bayesian parameter estimation in \texttt{cogwheel} with the \texttt{dynesty} sampler takes $\SI{4226}{\second}$ using one core. Our rapid PE framework, for the same injection, takes only $\SI{185}{\second}$ to run, providing a speed-up of a factor of $\sim 20$. 
	We note that it is possible to cut down the computational cost by not doing the maximum likelihood estimation on-the-fly. Instead, we can use the summary data from the search pipeline. 
	\begin{figure}[thb]
		\includegraphics[width=\columnwidth]{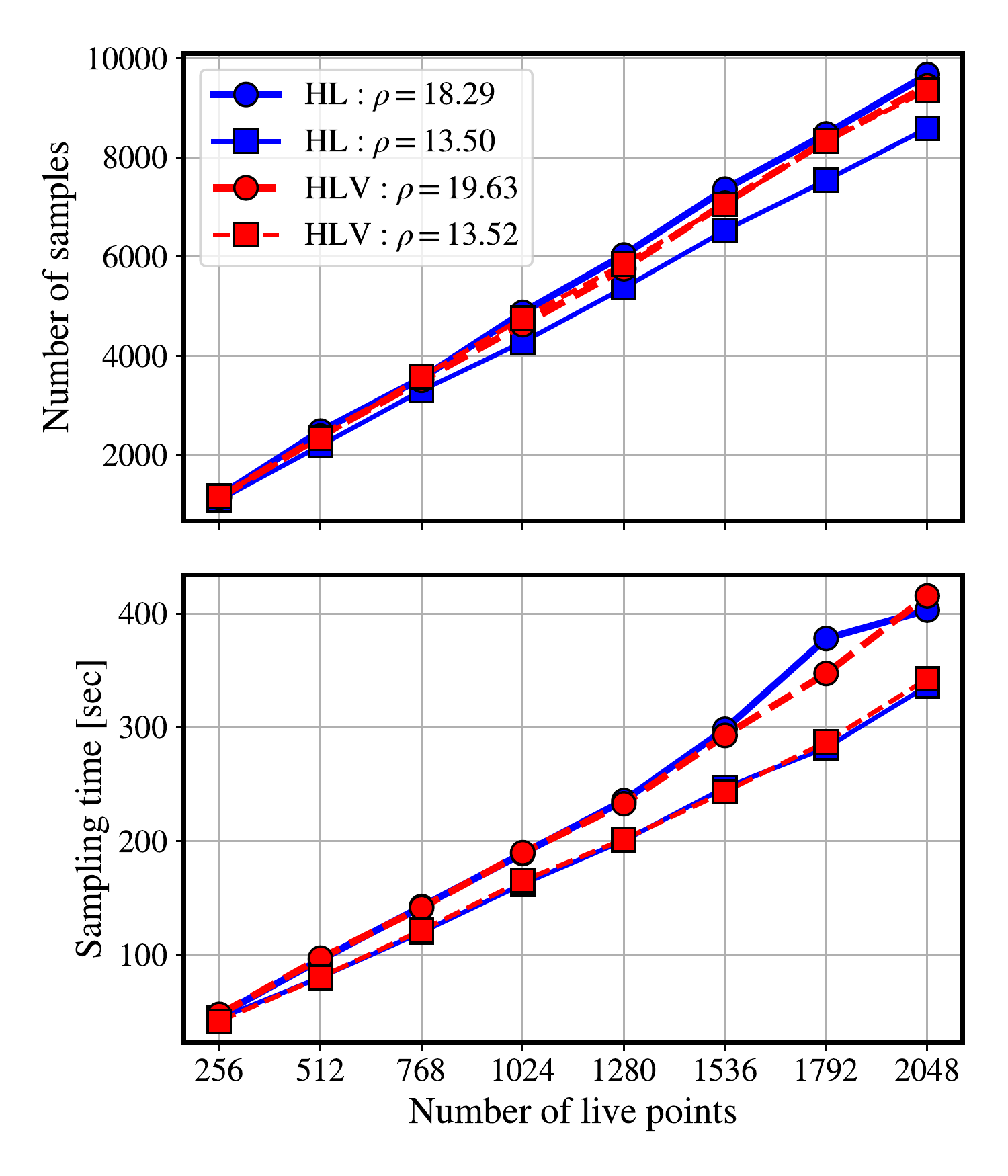}
		\caption{Number of samples in posteriors (upper panel) and sampling time (lower panel) as a function of the number of live points for the simulated BBH \texttt{Event-1} in both two detector (blue lines) and three detector (red lines) setup. We report results from two different
		values of the luminosity distances : $D=1000$ Mpc (circles) and $D=1500$ Mpc (squares).
		}
		\label{Fig:runtime}
	\end{figure}
	
	We also measure the time taken for likelihood evaluations in all cases. While standard Bayesian method in \texttt{cogwheel} with relative binning takes $\sim \SI{0.49}{m\second}$ for a single likelihood evaluation, factorized framework with relative binning takes $\sim\SI{3.04}{m\second}$. The likelihood evaluation cost in the factorized framework is higher than the standard Bayesian framework in \texttt{cogwheel} due to the marginalization over the extrinsic parameters. Our factorized framework is still faster than the standard framework as the sampler explores significantly reduced parameter space (only the masses and spins of the binary).
	
	We further note the time taken for sampling both the intrinsic and extrinsic parameters. In Fig. \ref{Fig:runtime}, we show the sampling time (lower panel) as well as the number of samples in posteriors (upper panel) as a function of the number of live points used in \texttt{PyMultiNest}. 
	We consider four different cases. We analyze the simulated BBH signal \texttt{Event-1} in both two-
	(LIGO-Hanford and LIGO-Livingston) and three-detector (LIGO-Hanford, LIGO-Livingston and Virgo) networks respectively. The signal has a signal-to-noise ratio (SNR) of 18.29 and 19.63 in two and three detector networks respectively. We then inject the same event at a larger distance of $D=1500$ Mpc yielding an SNR of 13.50 (13.52) in two-detector (three-detector) networks.
	We find that the sampling time and the number of samples generated increase almost linearly with the number of live points used. On average, our framework can generate $\sim 30$ samples per second. Furthermore, we find the sampling time to increase with an increase in signal-to-noise ratio.
	As the differences in the signal-to-noise ratios between the two-detector and three-detector networks analyses are small, the sampling time did not change significantly.
	
	We further demonstrate the efficacy of our rapid parameter estimation framework by analyzing \texttt{Event-2}, \texttt{Event-3} and \texttt{Event-4} whose masses, spins and distances are quite different from each other. We show the inferred posteriors for some of the key parameters (such as chirp mass, mass ratio, effective inspiral spin, distance and inclination angle) for all three events in Fig.~\ref{Fig:Event_2_3_4} (\texttt{Event-2} in blue, \texttt{Event-3} in orange and \texttt{Event-4} in green). Injected values are shown by vertical lines (for 1d posteriors) and dots (for 2d contours). In all three cases, the posteriors are consistent with the injected values. We also find that the runtime does not significantly depend on the injected binary source properties. Each run took $\sim \SI{200}{\second}$ to finish in one computing core.
	
	Finally, we repeat this analysis with a simulated $\sim \SI{128}{\second}$ long BNS signal injected in the LIGO-Virgo O3a noise data. We choose the detector frame total mass $M=1.83 M_{\odot}$, mass ratio $q=0.89$, spins $\chi_1=0.097$, $\chi_2=0.096$, luminosity distance $D=57.98$ Mpc and inclination angle $\iota=\pi/4$ and consider three-detector network. The signal has a network SNR of $\sim 18$ (similar to the simulated BBH \texttt{Event-1} in Table \ref{tab:Injections}). We find that the factorized parameter estimation framework is able to analyze the signal in $\sim \SI{250}{\second}$ with \texttt{IMRPhenomXAS} model (where the tidal deformability parameters are set to zero). If we use \texttt{IMRPhenomD\_NRTidalv2} and allow the sampler to explore the tidal parameters too apart from the masses and spins, runtime increases to $\sim \SI{650}{\second}$. This increase in runtime is solely due to the increase in dimensionality of the parameter space.
	
	\subsection{Analysis of GW170817}
	As a further evidence of the effectiveness of our method, we apply it to GW170817, the first GW signal 
	ever detected from a BNS system~\cite{LIGOScientific:2017vwq,LIGOScientific:2018hze}. GW170817 is also the loudest event detected so far with a signal-to-noise ratio of $\sim 33$, making it ideal to probe data analysis systematics. We use publicly available strain data released by the LVK collaboration \cite{LIGOScientific:2019lzm}.
    Furthermore, we adopt the same priors for all parameters as described in Sec.~\ref{subsubsec:prior}.
    Additionally, we impose uniform priors separately on the two tidal deformability parameters: $\Lambda_1, \Lambda_2 \in [0,5000]$. We find that the factorized parameter inference framework (with \texttt{IMRPhenomD\_NRTidalv2} waveform approximant and \texttt{PyMultiNest} sampler) takes only $\sim \SI{1124}{\second}$ in one core to generate posteriors for BNS source properties. In particular, the method takes $\sim \SI{133}{\second}$ to find the maximum likelihood while the rest of the time is mostly spent in sampling.
    
\begin{figure*}[thb]
		\includegraphics[scale=0.75]{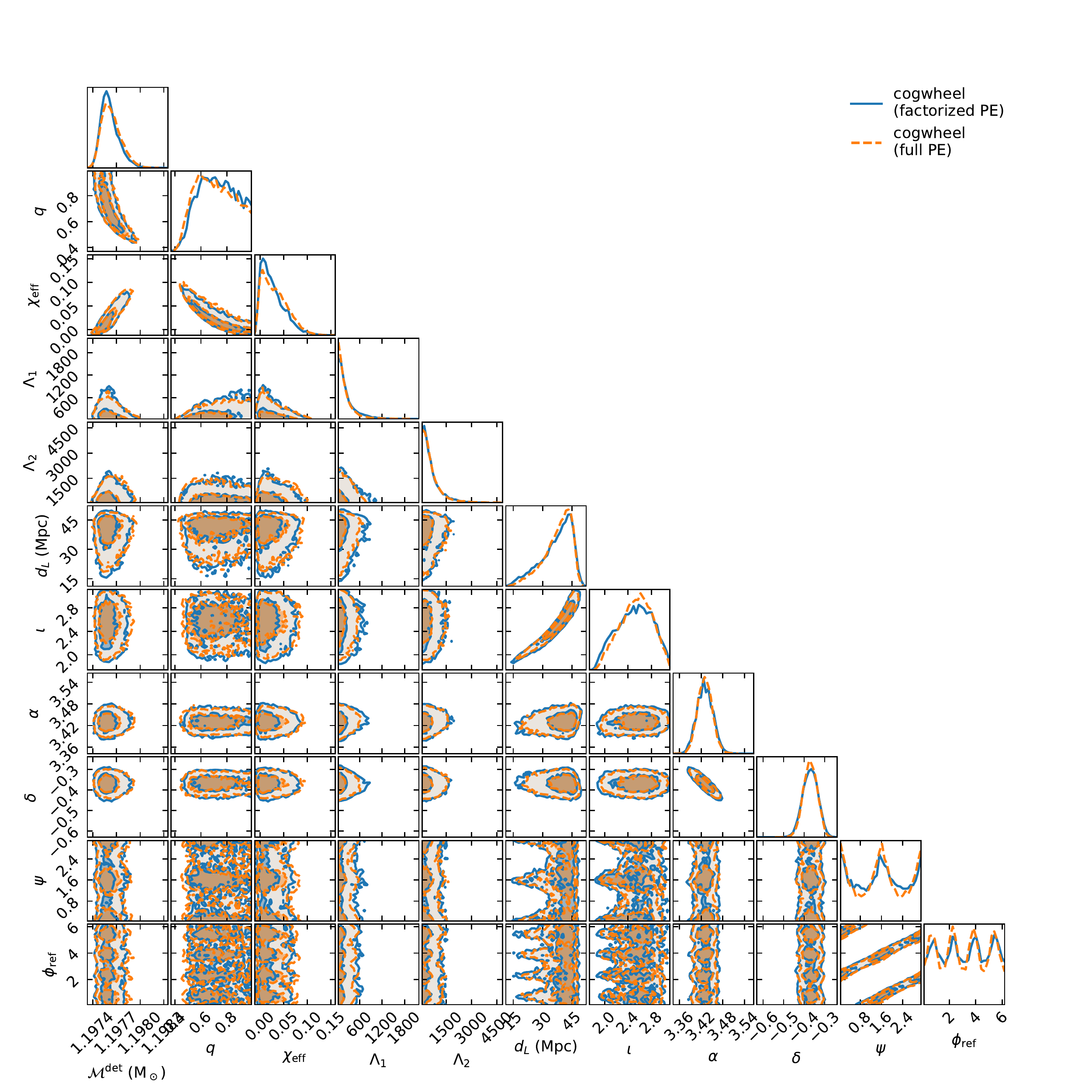}
		\caption{Posteriors obtained by analyzing \texttt{GW170817} using the factorized parameter estimation framework (blue) and standard Bayesian framework in \texttt{cogwheel} (orange) respectively. Contours enclose 50\% and 90\% of the posterior probability.
		}
		\label{Fig:GW17}
	\end{figure*}

    For comparison, we repeat the analysis using standard Bayesian parameter estimation in \texttt{cogwheel} keeping everything else the same. We show the posteriors from both the analyses in Fig \ref{Fig:GW17}. For all measured parameters, we find the posteriors from the factorized parameter estimation framework and standard \texttt{cogwheel} to be in good agreement. For all parameters, we quantify the difference between the one dimensional posteriors by computing JS divergence values. 
    We find that for all parameters JS divergence values are always less than 0.075 nats (with the smallest value of 0.032 nats for the orbital phase $\varphi$ and the highest value of 0.0749 nats for the effective inspiral spin $\chieff$) indicating a good match between the posteriors. 
    The standard Bayesian analysis, however, takes $\sim \SI{9.7}{\hour}$ in one core to complete. This translates to a speed-up of a factor of $\sim 50$ when factorized parameter estimation framework is used.
    
    This clearly demonstrates that the method is not only effective in rapid analysis of BBH signals, it also dramatically reduces the runtime for signals from aligned-spin BNS (and NSBH) mergers without any loss of accuracy.  
	
	\section{Discussion and Conclusion}
	\label{sec:discussion}
	In this paper, we have presented a fast, accurate parameter estimation within \texttt{cogwheel} for real-time GW inference for aligned-spin binaries. We use a semi-analytical marginalization over all seven extrinsic parameters and sample only the intrinsic parameters using Bayesian inference. We then use the sampled intrinsic parameter values to reconstruct the extrinsic parameters. Using an injected BBH signal, we have shown that the framework can accurately infer the binary properties. Furthermore, we have demonstrated that our result agrees well with a standard Bayesian inference analysis outcome. We notice a speed-up of $\sim 20$ compared to standard Bayesian inference analysis making our runtime to be $\sim \SI{200}{\second}$ (in one core). For BNSs, this framework takes $\sim \SI{250}{\second}$ to perform a parameter inference job (in one core). Our framework is further able to provide posteriors of all source parameters for GW170817, the first BNS event, in $\sim \SI{20} \minute$ as compared to $\SI{9}{\hour}$ for the standard Bayesian framework in \texttt{cogwheel}.

    While similar methods have previously been adopted in some parameter inference frameworks, more specifically in \texttt{RIFT} and in \texttt{bayestar}, there are significant differences in the implementation. In \texttt{RIFT}, intrinsic parameters are sampled in an iterative way in which, for each iterations, marginalized likelihood over the extrinsic parameters are computed numerically for a set of points across the parameter space. These marginalized likelihood data is then used in training a Gaussian Process Regression (GPR) fit from which all other likelihoods are drawn. For \texttt{bayestar}, the marginalization over the extrinsic parameters are done numerically using an adaptive grid. In contrast, we perform the marginalization using a combination of importance sampling over detector arrival times, Monte Carlo integration over inclination and polarization, and analytic integration over distance and orbital phase. Furthermore, while the primary objective of \texttt{bayestar} is to provide rapid localization for the GW transients, we aim to provide a rapid inference of all binary source properties including sky localization.
    
	Our method will be particularly useful for electromagnetic followups for BNS signals where robust real time sky localizations can be used for early warning to astronomers. 
	We note that while the current framework only works for aligned-spin systems with $(\ell,m)=(2,\pm2)$ modes, it can be generalized to precessing spin cases with higher harmonics. This will be one of the updates in future.
	
	\begin{acknowledgments}
    T.I. acknowledges support from NSF Grants No. PHY-2110496 and DMS-1912716. Part of this work is additionally supported by the Heising-Simons Foundation, the Simons Foundation, and NSF Grants Nos. PHY-1748958.
    Simulations were performed on CARNiE at the Center for Scientific Computing and Visualization Research (CSCVR) of UMassD, which is supported by the ONR/DURIP Grant No.\ N00014181255, the MIT Lincoln Labs {\em SuperCloud} GPU supercomputer supported by the Massachusetts Green High Performance Computing Center (MGHPCC) and ORNL SUMMIT under allocation AST166. 
    J.R. was supported by a grant from the Simons Foundation (216179, LB).
    T.V. acknowledges support by the National Science Foundation under Grant No. 2012086
    
    This research has made use of data or software obtained from the Gravitational Wave Open Science Center (gw-openscience.org), a service of LIGO Laboratory, the LIGO Scientific Collaboration, the Virgo Collaboration, and KAGRA. LIGO Laboratory and Advanced LIGO are funded by the United States National Science Foundation (NSF) as well as the Science and Technology Facilities Council (STFC) of the United Kingdom, the Max-Planck-Society (MPS), and the State of Niedersachsen/Germany for support of the construction of Advanced LIGO and construction and operation of the GEO600 detector. Additional support for Advanced LIGO was provided by the Australian Research Council. Virgo is funded, through the European Gravitational Observatory (EGO), by the French Centre National de Recherche Scientifique (CNRS), the Italian Istituto Nazionale di Fisica Nucleare (INFN) and the Dutch Nikhef, with contributions by institutions from Belgium, Germany, Greece, Hungary, Ireland, Japan, Monaco, Poland, Portugal, Spain. The construction and operation of KAGRA are funded by Ministry of Education, Culture, Sports, Science and Technology (MEXT), and Japan Society for the Promotion of Science (JSPS), National Research Foundation (NRF) and Ministry of Science and ICT (MSIT) in Korea, Academia Sinica (AS) and the Ministry of Science and Technology (MoST) in Taiwan. 
    
    A portion of this work was carried out while a subset of the authors were in residence
    at the Institute for Computational and Experimental Research in Mathematics (ICERM) in Providence, RI, during the Advances in Computational Relativity program. ICERM is supported by the National Science Foundation under Grant No. DMS-1439786. 

	\end{acknowledgments}  

	\bibliography{References}

\end{document}